\documentstyle[emulateapj,psfig,natbib,amsfonts,amsmath,graphicx]{article}
\def\ra#1#2#3{#1$^{\rm h}$#2$^{\rm m}$#3$^{\rm s}$}
\def\dec#1#2#3{$#1^\circ#2'#3''$}
\def\swift{{\it Swift}}
\def\nod{\nodata}
\def\grb{GRB\,050911}

\def\ociw{1}
\def\prince{2}
\def\hubble{3}

\begin{document}

\title{Galaxy Clusters Associated with Short GRBs. I. The Fields of
GRBs 050709, 050724, 050911 and 051221a}

\author{
E.~Berger\altaffilmark{\ociw,}\altaffilmark{\prince,}\altaffilmark{\hubble},
M.-S.~Shin\altaffilmark{\prince},
J.~S.~Mulchaey\altaffilmark{\ociw},
and T.~E.~Jeltema\altaffilmark{\ociw}
}

\altaffiltext{\ociw}{Observatories of the Carnegie Institution
of Washington, 813 Santa Barbara Street, Pasadena, CA 91101}

\altaffiltext{\prince}{Princeton University Observatory,
Peyton Hall, Ivy Lane, Princeton, NJ 08544}

\altaffiltext{\hubble}{Hubble Fellow}

\begin{abstract} 
We present a search for galaxy clusters in the fields of three
bona-fide short GRBs (050709, 050724, and 051221a) and the putative
short burst GRB\,050911 using multi-slit optical spectroscopy.  These
observations are part of a long-term program to constrain the
progenitor age distribution based on the fraction of short GRBs in
galaxy clusters and early-type galaxies.  We find no evidence for
cluster associations at the redshifts of the first three bursts, but
we confirm the presence of the cluster EDCC 493 within the error
circle of GRB\,050911 and determine its redshift, $z=0.1646$, and
velocity dispersion, $\sigma\approx 660$ km s$^{-1}$.  In addition,
our analysis of \swift/XRT observations of this burst reveals diffuse
X-ray emission coincident with the optical cluster position, with a
luminosity, $L_X\approx 4.9\times 10^{42}$ erg s$^{-1}$, and a
temperature, $kT\approx 0.9$ keV.  The inferred mass of the cluster is
$2.5\times 10^{13}$ M$_\odot$, and the probability of chance
coincidence is about $0.1-1\%$, indicating an association with \grb\
at the 2.6-3.2$\sigma$ confidence level.  A search for diffuse X-ray
emission in coincidence with the fifteen other short GRBs observed
with XRT and {\it Chandra} reveals that with the exception of the
previously-noted cluster ZwCl 1234.0+02916 likely associated with
GRB\,050509b, no additional associations are evident to a typical
limit of $3\times 10^{-14}$ erg s$^{-1}$ cm$^{-2}$, or $M\lesssim
5\times 10^{13}$ M$_\odot$ assuming a typical $z=0.3$.  The resulting
fraction of short GRBs hosted by galaxy clusters of about $20\%$ is in
rough agreement with the fraction of stellar mass in clusters of $\sim
10-20\%$.
\end{abstract}
 

\section{Introduction}
\label{sec:intro}

Several lines of evidence indicate that the progenitors of
short-duration, hard-spectrum gamma-ray bursts (GRBs) are related to
an old stellar population.  These include the localization of
GRB\,050724 \citep{bpc+05,pbc+06}, and most likely GRB\,050509b
\citep{gso+05,bpp+06}, to bright elliptical galaxies, the lack of
supernova emission in several low redshift short GRBs
\citep{hsg+05,ffp+05,bpp+06,sbk+06}, and the location of GRB\,050709
outside of any star forming region in its host galaxy \citep{ffp+05}.
While this supports the popular notion that the progenitors are
compact objects binaries (DNS or NS-BH; e.g.,
\citealt{elp+89,npp92,rrd03}), the lack of direct observations (e.g.,
gravitational waves or a sub-relativistic, radioactive component:
\citealt{lp98,kul05}), suggests that a more detailed understanding of
the progenitor population has to rely on statistical studies.

In this vein, \citet{ngf05} and \citet{gp06} argue that the redshift 
and luminosity distributions of \swift\ and BATSE short GRBs, are 
inconsistent with the nominal merger time distribution of DNS binaries 
in the Milky Way, $P(\tau)\propto \tau^{-1}$ \citep{clm+04}.  They 
further conclude that the typical age of the progenitors is old, 
$\gtrsim 4$ Gyr \citep{ngf05}.  Similarly, \citet{gno+05} and 
\citet{zr06} propose that the relative fractions of short GRBs in 
early- and late-type galaxies should constrain the progenitor lifetime 
distribution.  This test derives from the fact that, on average, stars 
in early-type galaxies form earlier than in late-type galaxies.

An an extension of this idea, the preponderance of short GRBs in
galaxy clusters can also be used to constrain the age distribution and
nature of the progenitor population.  In the framework of $\Lambda$CDM
cosmology, high-resolution numerical simulations suggest that the
oldest stars reside in dense galaxy cluster environments, typically
within $\sim 150$ kpc of the cluster center \citep{ws00,dsw+06}.  This
is supported by observations, which indicate that the difference in
formation epochs for stars in cluster and field ellipticals is $\sim
1-3$ Gyr (e.g., \citealt{brc+98,ksc+02,tmb+05}).  Since $M$* in
clusters is larger than in the field (e.g., \citealt{bbb+06}), this
leads to an even more pronounced difference in star formation history
than just the difference between early- and late-type galaxies
discussed in \citet{zr06}.

In addition, the early-type fraction in clusters, $\sim 60\%$, is
twice as high as in the field \citep{dre80,wgj93}, suggesting that the
fraction of short GRBs in clusters is intimately related to their rate
in early-type galaxies.  This is of particular importance in cases
where the positional accuracy of the burst is not sufficient to
associate it with a particular galaxy, but may be sufficient to
associate it with a galaxy cluster.  Finally, the specific frequency
of globular clusters is at least a factor of a few higher in bright
cluster ellipticals than in field galaxies \citep{har91}.  Thus, the
fraction of short bursts in galaxy clusters may shed light on whether
globular clusters are an efficient site for the formation of short GRB
progenitors, as proposed by \citet{gpm06}.

A complete search for clusters is also important from an observational 
point of view.  These associations can be made at high significance 
based on the prompt $\gamma$-ray positions alone (typically, $2-3'$), 
whereas associations with individual galaxies require accurate 
positions ($\sim 1-10''$) from the optical, radio, or X-ray afterglow.  
Since the afterglow brightness correlates with the circumburst 
density\footnotemark\footnotetext{The X-ray luminosity is usually 
considered to be insensitive to the circumburst density ($n$), but 
this is only true when the X-ray band is located above the synchrotron 
cooling frequency, requiring a relatively high density ($n\gtrsim 
10^{-3}$ cm$^{-3}$) to begin with.  At lower densities, typical of the 
intracluster medium, the afterglow luminosity in all bands is expected 
to be low.} the latter approach may produce an observational bias in 
favor of gas-rich or disk galaxies.

To date, three associations of short GRBs with galaxy clusters have
been claimed.  GRB\,050509b appears to reside in the cluster ZwCl
1234.0+02916 at $z=0.226$ \citep{peh+05,bpp+06}, GRB\,050813 is
apparently associated with a cluster at $z\sim 1.8$ (\citealt{ber06};
Gladders et al. in prep.), and GRB\,790613 may be associated with the
cluster Abell 1892 at $z\approx 0.09$ \citep{gno+05}.  The statistical
significance of these associations is $\sim 3\sigma$.

Motivated by these considerations we began the first systematic search
for galaxy clusters hosting short GRBs, using multi-slit optical
spectroscopy and archival X-ray observations.  This is part of a
long-term program to constrain the age distribution of the progenitors
using the properties of their large-scale environments.  Here we
present spectroscopy in the fields of GRBs 050709, 050724, 050911, and
051221a.  We also re-analyze all of the available X-ray observations
of short GRB fields to search for diffuse emission from hot
intracluster gas associated with potential clusters.  The layout of
the paper is as follows.  The optical observations are described in
\S\ref{sec:spec}, and the X-ray analysis in presented in
\S\ref{sec:xray}.  In \S\ref{sec:res} we summarize the results of our
search, including a determination of the optical and X-ray properties
of the cluster EDCC 493, which coincides with GRB\,050911.  We draw
initial conclusions in \S\ref{sec:disc}.  Throughout the paper we use
the standard $\Lambda$CDM cosmology with $H_0=71$ km s$^{-1}$
Mpc$^{-1}$, $\Omega_m=0.27$, and $\Omega_\Lambda=0.73$.

\section{Observations}
\label{sec:obs}

For a detailed discussion of the four bursts studied in this paper we
refer the reader to the following publications.  GRB\,050709:
\citet{ffp+05}, \citet{hwf+05}, and \citet{vlr+05}.  GRB\,050724:
\citet{bpc+05}, \citet{ctl+06}, and \citet{bcb+05}.  GRB\,050911:
\citet{pkl+06}.  GRB\,051221a: \citet{sbk+06} and \citet{bgc+06}.  We
note that the classification of \grb\ is somewhat ambiguous.
Formally, $T_{90}$ for this burst is $\sim 16$ s \citep{pkl+06}, but
the light curve is dominated by an initial pair of short pulses with a
total duration of about 1.5 s, followed by a softer component with a
slow rise and decay.  The latter component may be similar to the soft
tails observed in GRBs 050709 \citep{vlr+05} and 050724
\citep{bcb+05}, and moreover would have been missed by BATSE
\citep{pkl+06}.  In this framework, \grb\ would be classified as a
short GRB.

\subsection{Optical Spectroscopy}
\label{sec:spec}

We selected targets for spectroscopy based on imaging observations
from the {\it Hubble Space Telescope} (050709), the Magellan/Clay Low
Dispersion Survey Spectrograph (050724), the Gemini Multi-Object
Spectrograph (051221a), and the du Pont 100-inch telescope at Las
Campanas Observatory (050911).  We used the program SExtractor
\citep{ba96} to estimate source magnitudes and to separate stars and
galaxies (using a stellarity index of $<0.5$ for galaxies).  From the
spectroscopy we find a star interloper fraction of $5\%$ for
GRB\,050911, $12\%$ for GRB\,051221a, and zero for GRBs 050724 and
050709.  The objects range in brightness from $R=17.9$ to $21.9$ mag
(050709; $R_{\rm host}\approx 21.2$ mag), $I=17.6$ to $21.1$ mag
(050724; $I_{\rm host}\approx 18.6$ mag), $R=15.3$ to $19.7$ mag
(050911), and $r'=21$ to $22.5$ mag (051221a; $r_{\rm host}\approx
22.0$ mag).  The final source catalogs (not including stars) contain
33 objects (050709), 21 objects (050724), 79 objects (051221a), and 38
objects (051109).

All spectra were obtained with the Low Dispersion Survey Spectrograph 
(LDSS3) mounted on the Magellan/Clay 6.5-m telescope using a 300 lines 
mm$^{-1}$ grism, which provides a resolution of about $6$ \AA.  For 
GRB\,050724 we also used a volume-phase holographic grism, which 
provides a resolution of about 2 \AA.  The log of the observations is 
provided in Table~\ref{tab:obs}.  We reduced the data with the {\tt 
cosmos} software 
package\footnote{http://shimura.ociw.edu:8200/Code/Groups/Cosmos}, 
using flat and HeNeAr arc exposures obtained following each mask 
exposure.  The data were bias-subtracted, flat-fielded with a 
response-corrected flat, and sky-subtracted using a 2-d spline fit.  
The spectra were then extracted and combined following cosmic-ray 
rejection.  Redshifts were determined manually using the IRAF task 
{\tt splot} to measure the absorption and/or emission line positions.  
We obtained redshifts for a total of 151 galaxies in the four GRB 
fields, or an overall success rate of $79\%$.

\subsection{X-ray Data}
\label{sec:xray}

We retrieved from the High Energy Astrophysics Science Archive
Research Center\footnote{http://heasarc.gsfc.nasa.gov/W3Browse/} all
publicly available observations of short GRBs taken with the \swift\
X-ray telescope (XRT) and the {\it Chandra X-ray Observatory}.  A
summary of the observations and exposure times for the sixteen
available GRBs is given in Table~\ref{tab:xray}.

We processed the XRT data with the {\tt xrtpipeline} script packaged
within the HEAsoft software, using the default grade selection and
screening parameters.  For the {\it Chandra} data we used the {\tt
evt2} files provided by HEASARC.  All event files were further
filtered for the energy range $0.5-7$ keV using {\tt xselect}.  We
searched for diffuse emission at the positions of the short GRBs
visually using the CIAO routine {\tt csmooth} to construct smoothed
images.

\section{Results}
\label{sec:res}

\begin{figure}
\epsscale{1}
\centerline{\psfig{file=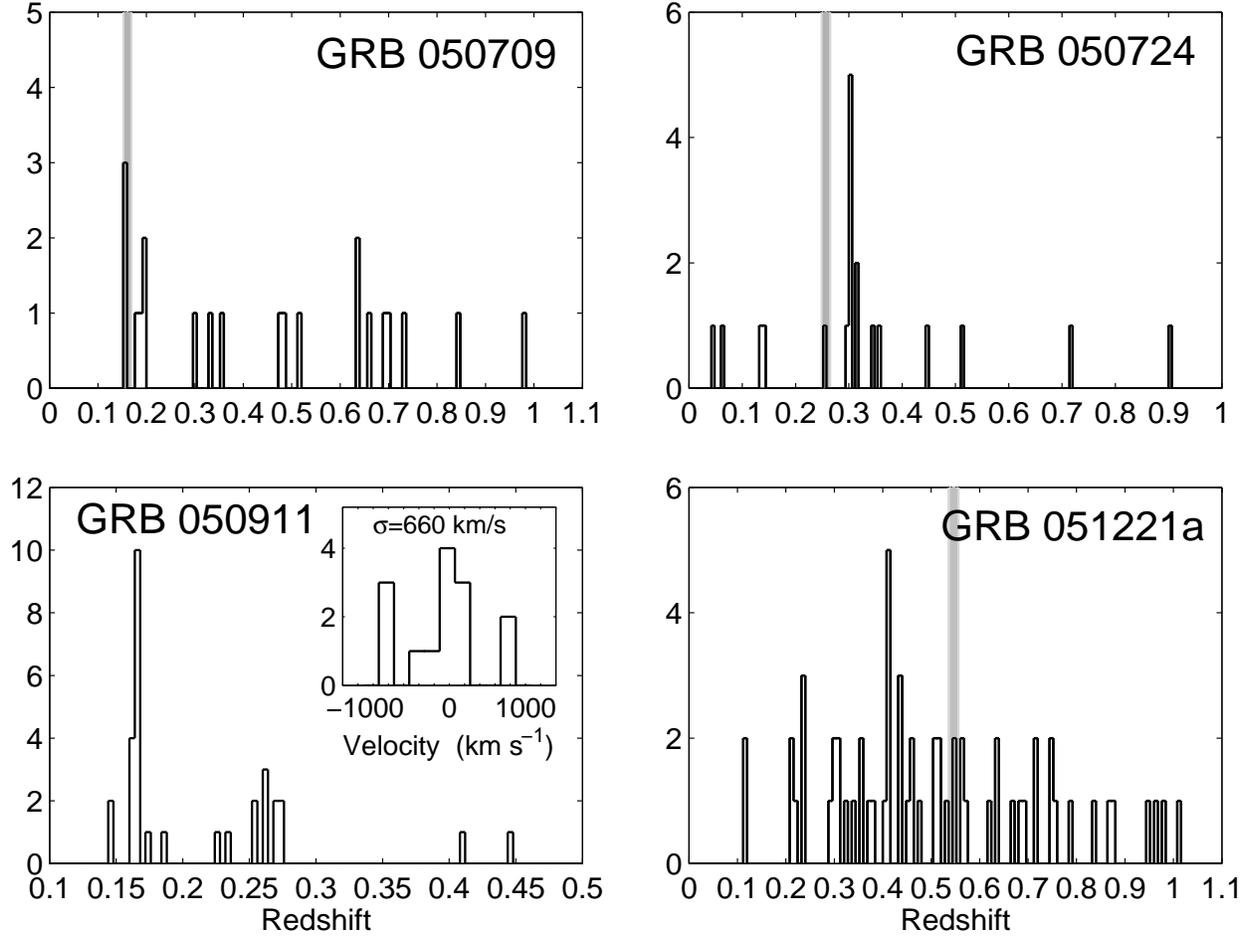,width=6.5in}}
\caption{Redshift distributions for galaxies observed in the fields of
GRBs 050709, 050724, 050911, and 051221a.  The gray stripes designate
a range of $\pm 2000$ km s$^{-1}$ (dark) and $\pm 4000$ km s$^{-1}$
(light) centered on the redshift of the burst.  In the case of
GRB\,050709 there are two galaxies in the field within $2000$ km
s$^{-1}$ of the burst redshift ($z=0.161$), at $z=0.154$ and
$z=0.156$.  In the field of GRB\,050724 we do not find any galaxies
within $9000$ km s$^{-1}$ of the burst redshift ($z=0.257$).  Two
galaxies in the field of GRB\,051221a are located within $2000$ km
s$^{-1}$ of the burst redshift ($z=0.5465$), at $z=0.550$ and
$z=0.544$.  Finally, the cluster EDCC 493 at $z=0.1646$ is clearly
visible in the field of \grb.  The velocity dispersion of this cluster
is $660$ km s$^{-1}$ (inset).
\label{fig:grbsz}}
\end{figure}

We show the results of our spectroscopic observations overlaid on
images of each of the four fields in
Figures~\ref{fig:050709}--\ref{fig:051221a}.  The redshift
distributions are presented in Figure~\ref{fig:grbsz}.  For
GRB\,050724 we do not find any galaxies in the $8'$ diameter field,
within $\sim 9000$ km s$^{-1}$ of the host redshift of $z=0.257$.  A
possible background galaxy group or cluster is located at $z\approx
0.3$, but we do not detect any coincident X-ray emission with a limit
of $L_X\lesssim 8.3\times 10^{42}$ erg s$^{-1}$ (at $z=0.3$).  This
indicates that this background structure is at most a poor cluster.

Similarly, of the 21 galaxies with spectroscopic redshifts in the
$2'\times 2'$ field of GRB\,050709, we find only two within $2000$ km
s$^{-1}$ of the burst redshift, $z=0.161$.  This is unlikely to
constitute a significant structure, and in fact we place a limit of
$L_X\lesssim 1.6\times 10^{42}$ erg s$^{-1}$ on diffuse X-ray emission
associated with the burst environment.

In the field of GRB\,051221a we find a nearly uniform redshift
distribution between $z\sim 0.1$ and $1$, with only two galaxies in
the $5'\times 5'$ field located within 2000 km s$^{-1}$ of the burst
redshift, $z=0.5465$.  However, these galaxies, at $z=0.550$ and
$z=0.544$, are situated about $3.0'$ (1.2 Mpc) and $2.8'$ (1.1 Mpc)
away from the GRB host galaxy, respectively, suggesting that this is
not likely to be a significantly overdense structure.  The limit on
diffuse X-ray emission coincident with the burst position is
$L_X\lesssim 1.7\times 10^{43}$ erg s$^{-1}$, or about $50\%$ fainter
than the X-ray luminosity of the cluster associated with GRB\,050509b
(\citealt{peh+05,bpp+06}; Table~\ref{tab:xray}).

\subsection{The Galaxy Cluster EDCC 493 in the Error Circle of \grb}
\label{sec:edcc}

The BAT error circle of \grb, centered on $\alpha=$\ra{00}{54}{52.4},
$\delta=$\dec{-38}{51}{42.8} (J2000) with an uncertainty of $2.8'$
radius \citep{pkl+06}, intersects the galaxy cluster EDCC 493
\citep{ber05}.  This cluster has an Abell radius of about $9.5'$
\citep{lnc+92}.

Our spectroscopic observations in this field quantify the properties
of the cluster.  We obtain redshifts for fourteen cluster members,
including the brightest elliptical galaxy.  The properties of the
cluster galaxies are summarized in Table~\ref{fig:050911}.  We
estimate the cluster velocity dispersion using the {\tt ROSTAT}
package \citep{bfg90}.  We consider all galaxies within $3000$ km
s$^{-1}$ of the mean cluster redshift, and calculate the biweight
estimators of the location (mean velocity) and scale (velocity
dispersion).  Objects with velocities greater than three times the
velocity dispersion are removed from the sample and a new location and
scale are calculated iteratively until no more objects are clipped. In
this particular system, no objects were clipped from the original
list.  We find $\sigma=660^{+135}_{-95}$ km s$^{-1}$ and a redshift of
$z=0.1646$ (Figure~\ref{fig:grbsz}).

The early-type fraction in our sample of fourteen cluster members is
$80\pm 25\%$, at the high end of the distribution for groups/clusters
with a similar velocity dispersion \citep{mlf+06}.  This suggests that
\grb\ was most likely associated with an early-type galaxy.  Another
potential implication is that EDCC 493 is more evolved than a typical
cluster of the same mass, perhaps a reflection of the old age of the
GRB progenitor system.  The determination of the early-type fraction
for a larger sample of cluster members will show whether this effect
is real.

\begin{figure}
\epsscale{1}
\includegraphics[angle=270,width=7in]{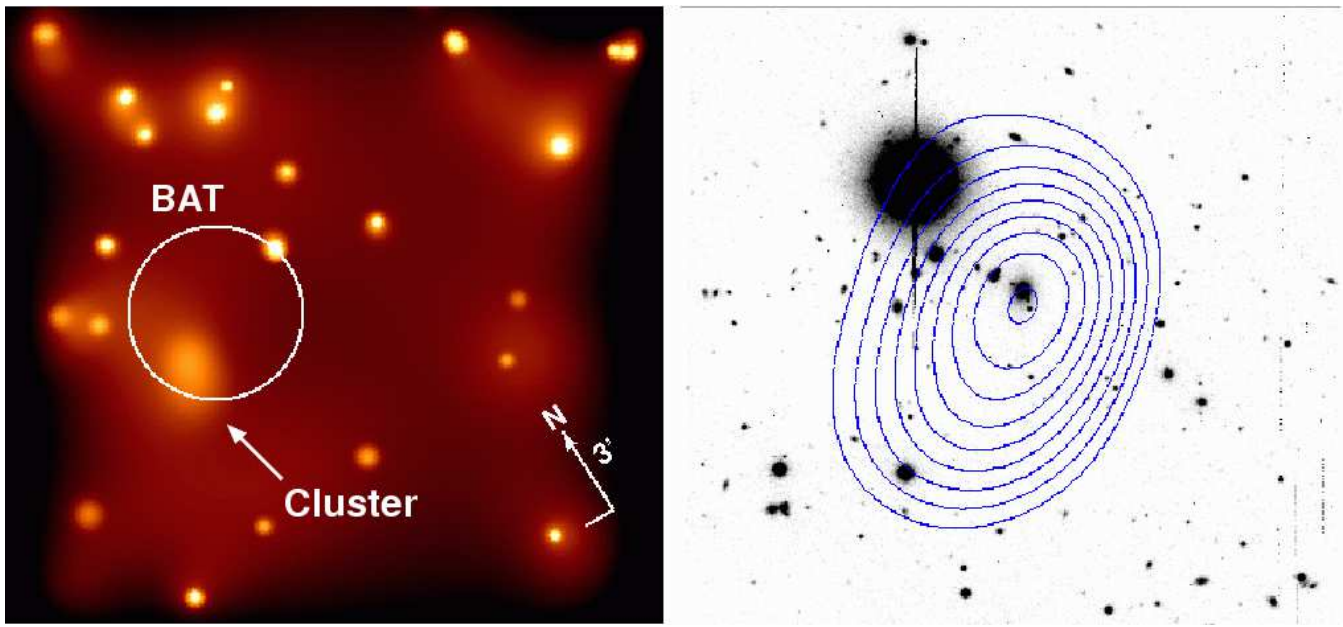}   
\caption{{\it Left:} Smoothed \swift/XRT image of the field of \grb.
The diffuse X-ray emission from EDCC 493 is clearly visible in the
southeast corner of the BAT error circle.  {\it Right:} X-ray contours
overlaid on the optical $I$-band image indicating that the diffuse
X-ray emission is centered on the brightest cluster galaxy (with
$L\sim 3\,L$*; see \S\ref{sec:edcc}).
\label{fig:050911xray}}
\end{figure}

We also detect diffuse X-ray emission in the XRT data coincident with
the optical cluster position, at $\alpha=$\ra{00}{55}{01.39},
$\delta=$\dec{-38}{52}{45.8} (J2000), with an uncertainty of about
$5''$ in each coordinate (Figure~\ref{fig:050911xray}).  This position
is $8.5''$ west and $15.7''$ south of the optical position of the
bright cluster elliptical.

\begin{figure}
\epsscale{1}
\includegraphics[angle=270,width=7.5in]{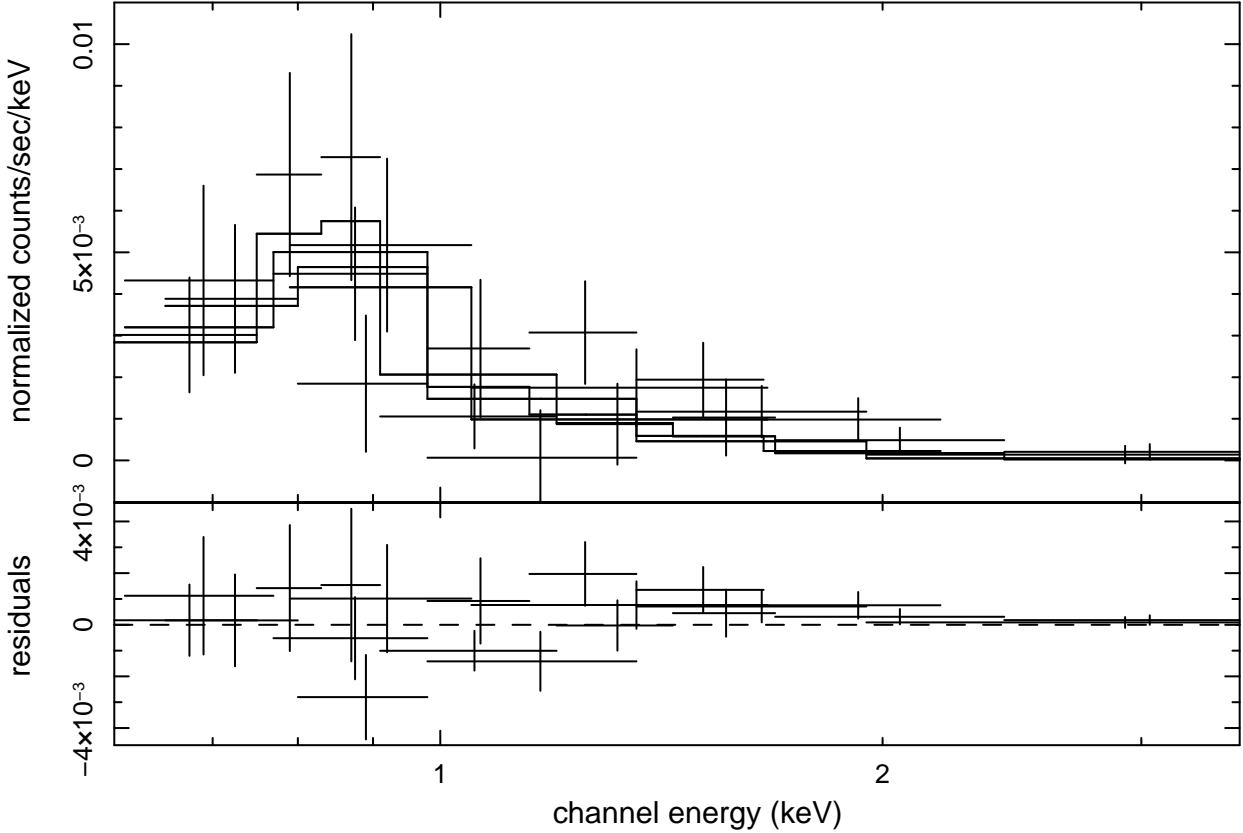}   
\caption{Spectrum of the diffuse X-ray emission from the cluster EDCC
493 located in the error circle of \grb.  We simultaneously fit the
counts extracted from all four XRT observations of this field
(Table~\ref{tab:xray}).  The data are binned with at least ten counts
per bin.  We use a MEKAL model fit with an abundance fixed at $0.3$
Z$_\odot$ and an absorbing column density of $N_H=2.7\times 10^{20}$
cm$^{-2}$ cm$^{-2}$ \citep{dl90}.  The best-fit parameters are
$kT=0.9^{+0.3}_{-0.2}$ keV, and $L_X=4.9^{+1.3}_{-1.2}\times 10^{42}$
erg s$^{-1}$ ($\chi^2_r=1.0$ for 19 degrees of freedom).
\label{fig:g050911spec}}
\end{figure}

To determine the source X-ray properties we extract counts for each
individual observation in an elliptical aperture with semi-major and
semi-minor axes of $120''$ and $80''$, respectively, selected to match
the scale at which the cluster diffuse emission matches the background
level.  We then bin the extracted counts in energy such that each bin
contains at least ten counts.  Using a MEKAL model fit to the energy
range of $0.5-7$ keV with an abundance fixed at $0.3$ Z$_\odot$
\citep{ml97} and an absorbing column density of $N_H=2.7\times
10^{20}$ cm$^{-2}$ cm$^{-2}$ \citep{dl90}, we find
$kT=0.9^{+0.3}_{-0.2}$ keV, and $L_X=4.9^{+1.3}_{-1.2}\times 10^{42}$
erg s$^{-1}$ ($\chi^2_r=1.0$ for 19 degrees of freedom).  The data and
model fit are shown in Figure~\ref{fig:g050911spec}.

The measured velocity dispersion and X-ray luminosity are in good
agreement with values measured for galaxy groups and poor clusters
from the ROSAT Deep Cluster Survey \citep{mlf+06}, but the temperature
is somewhat lower than expected in comparison to the compilation of
\citet{hms99} from which we estimate $kT\sim 2.5$ keV.  Using the
cluster mass-temperature relation \citep{app05} we estimate a mass of
$M_{500}\approx 2.5\times 10^{13}$ M$_\odot$, or about a factor of six
times lower than the cluster ZwCl 1234.0+02916 in the field of
GRB\,050509b (see \S\ref{sec:xdata}).

We next assess the probability that \grb\ is associated with EDCC 493.
From the ${\rm log}\,N-{\rm log}\,S$ relation for X-ray clusters in
the ROSAT Deep Cluster Survey we find that for a flux of $F_X>7.7
\times 10^{-14}$ erg $^{-1}$ cm$^{-2}$ the surface density of sources
is about 1.4 deg$^{-2}$ \citep{rcn+98}.  The probability of chance
coincidence with the $2.8'$ radius error circle is therefore
$9.6\times 10^{-3}$.  Thus, we conclude that the association between
\grb\ and EDCC 493 is significant at the $2.6\sigma$ confidence level.
We note that taking into account the thirteen additional searches
(excluding GRB\,050509b; see \S\ref{sec:xdata}) the probability of
finding such a cluster in any of the BAT error circles is about $12\%$
(or, $1.6\sigma$).  Of course, of the other fourteen short bursts,
eleven have much more accurate positions from X-ray, optical, and/or
radio afterglow observations.

If we consider the X-ray luminosity of the cluster, we find that the
volume density of clusters with $L_X=4.9\times 10^{42}$ erg s$^{-1}$
is ${\rm d}N/{\rm d}L_X\approx 3.3\times 10^{-5}$ Mpc$^{-3}$
($10^{44}$ erg s$^{-1}$)$^{-1}$ \citep{rcn+98}.  Integrating the X-ray
luminosity function with $\alpha=-1.83$ \citep{rcn+98} we find
$N(L>L_{\rm X,EDCC 493})\approx 3.2\times 10^{-6}$ Mpc$^{-3}$, or
$2.7\times 10^{-5}$ arcmin$^{-2}$ within the distance to EDCC 493.
The probability of finding such a cluster within the error circle is
therefore $7\times 10^{-4}$, or a $3.4\sigma$ significance level for
an association with \grb.

Turning to the brightest cluster galaxy (BCG), we find from the 2MASS
catalog that it has $K=13.5\pm 0.1$ mag in a $6.6''$ aperture, or
$I-K=4.1\pm 0.1$ mag compared to our $I$-band photometry.  The surface
density of sources with equal or greater brightness is about 18
deg$^{-2}$, or a probability of $0.12$ of finding such an object
within the error circle.  The rest-frame $K$-band absolute magnitude
of the BCG is $M_K\approx -25.4$ mag, or $L\approx 3\,L$* in
comparison to the luminosity function from the 2dF Galaxy Redshift
Survey and 2MASS \citep{cnb+01}.  Integrating the $K$-band luminosity
function we find that the number density of such galaxies is
$N(L>3L^*)\approx 5.1\times 10^{-5}$ Mpc$^{-3}$, or $4.3\times
10^{-4}$ arcmin$^{-2}$ within the distance to EDCC 493.  Thus, the
probability to find such a luminous galaxy within the BAT error circle
of \grb\ is about 0.01 ($2.6\sigma$ confidence level).

We therefore conclude that the probability of chance association is
only 0.1-1$\%$, and it is therefore likely that \grb\ occurred within
the cluster.  At the redshift of EDCC 493, the isotropic-equivalent
$\gamma$-ray energy release of the burst was $1.9\times 10^{49}$ erg,
similar to that of other short GRBs \citep{sbk+06}.

\subsection{Diffuse X-ray Emission}
\label{sec:xdata}

\begin{figure}
\epsscale{1}
\includegraphics[angle=270,width=7in]{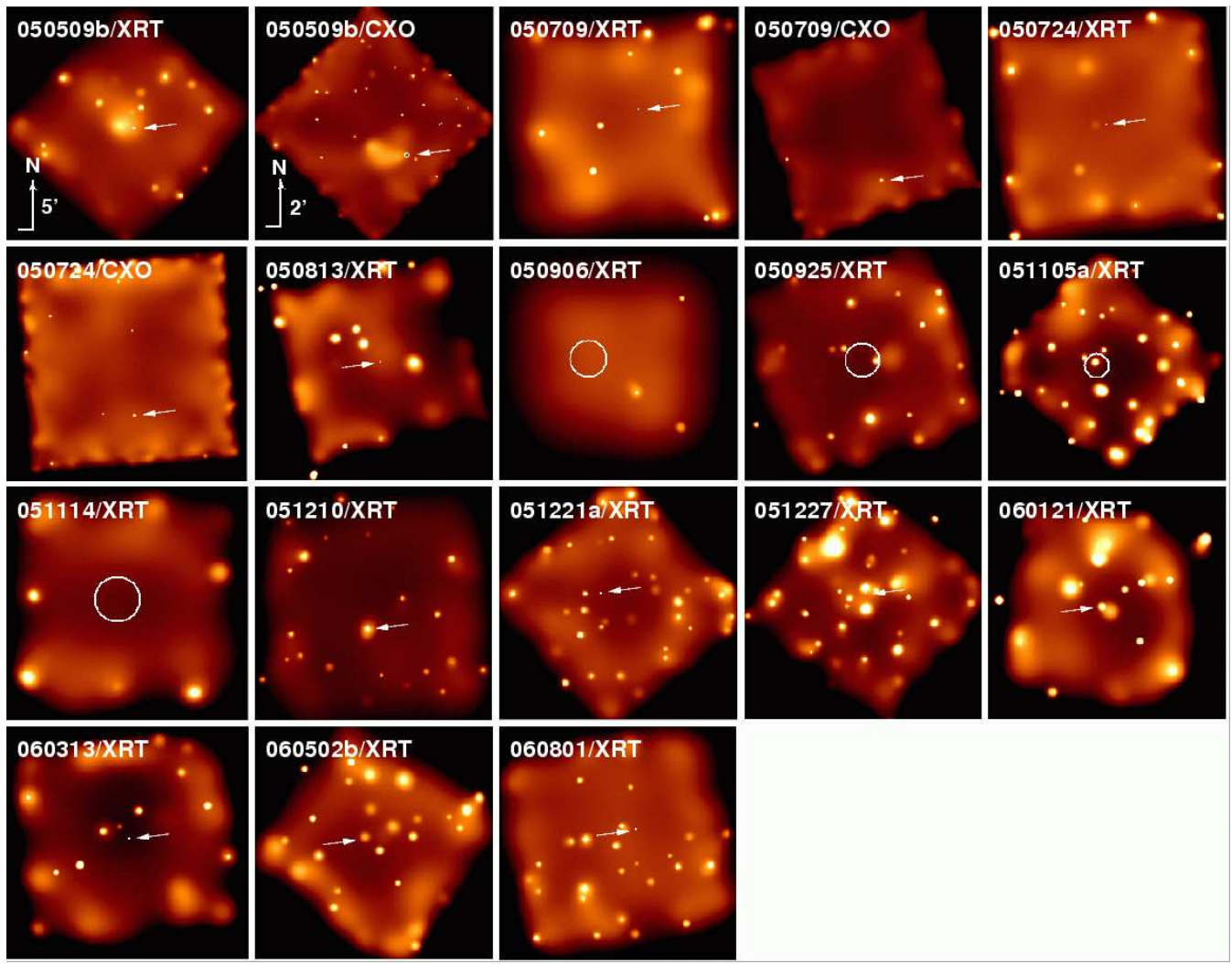}   
\caption{Smoothed \swift/XRT and {\it Chandra} X-ray images of the
fields of fifteen short GBRs.  The location of the burst is marked
with a circle when only a BAT position is available, or an arrow if an
afterglow has been detected.  The diffuse X-ray emission from the
cluster ZwCl 1234.0+02916 coincident with GRB\,050509b
\citep{peh+05,bpp+06} is clearly visible in the XRT and {\it Chandra}
images.
\label{fig:xray}}
\end{figure}

In Figure~\ref{fig:xray} we show smoothed XRT and {\it Chandra} X-ray
images of the fields of the fifteen short GRBs.  With the exception of
the previously detected diffuse X-ray emission from the cluster ZwCl
1234.0+02916 coincident with GRB\,050509b \citep{peh+05,bpp+06}, we do
not detect clear diffuse X-ray emission in coincidence with any of
other short GRBs.  We calculate upper limits on the X-ray flux using
$3\times\sqrt{N_{\rm cts}}/t_{\rm exp}$ within a $2'$ radius circle
centered on each GRB position, where $N_{\rm cts}$ is the total number
of counts within the circle in the $0.5-7$ keV range (effectively the
background level), and $t_{\rm exp}$ is the total exposure time.  To
convert from count rate to flux we assume a thermal bremsstrahlung
model with $kT=1$ keV and an absorbing column given by the
\citet{dl90} value for each burst.  The typical upper limits are
$\lesssim 3\times 10^{-14}$ erg s$^{-1}$ cm$^{-2}$, or about a factor
two lower than for EDCC 493.  Assuming a typical redshift, $z\sim
0.3$, the corresponding luminosity limit is $\lesssim 7\times 10^{42}$
erg s$^{-1}$, or roughly $M_{500}\lesssim 5\times 10^{13}$ M$_\odot$.

For the cluster ZwCl 1234.0+02916 associated with GRB\,050509b we
derive a temperature of $kT=3.0^{+0.9}_{-0.6}$ keV and an unabsorbed
luminosity of $L_X=4.5^{+0.7}_{-0.6}\times 10^{43}$ erg s$^{-1}$
($\chi^2_r=0.64$ for 12 degrees of freedom).  The absorbing column
density is $N_H=1.1^{+0.6}_{-0.5} \times 10^{21}$ cm$^{-2}$, about an
order of magnitude larger than the \citet{dl90} value for the Galactic
column.  Here we used a MEKAL model with the abundance fixed to $0.3$
Z$_\odot$.  We note that our derived temperature is in good agreement
with the value of about 3.65 keV found by \citet{peh+05}, but is lower
than the value of 5.25 keV found by \citet{bpp+06}.  Our derived
luminosity is about $50\%$ higher than that of \citet{peh+05}.  The
inferred cluster mass is $R_{500}=1.6^{+1.1}_{-0.6}\times 10^{14}$
M$_\odot$.

\section{Discussion}
\label{sec:disc}

We present the first systematic search for galaxy clusters hosting
short GRBs using multi-slit optical spectroscopy in the fields of GRBs
050709, 050724, 050911, and 051221a, and a re-analysis of all publicly
available X-ray observations.  Future papers in this series will
present optical spectroscopy of additional short GRB fields and a
detailed analysis of the galaxy cluster statistics.  No apparent
clusters are found in the fields of GRBs 050709, 050724, and 051221a
from optical and X-ray observations.  In the error circle of the
putative short burst \grb\ we show that the cluster EDCC 493 has a
mean redshift, $z=0.1646$, a velocity dispersion, $\sigma=660$ km
s$^{-1}$, an X-ray temperature, $kT=0.9$ keV, and a luminosity,
$L_X=4.9\times 10^{42}$ erg s$^{-1}$.  These values are typical for
poor clusters, and the inferred mass is about $2.5\times 10^{13}$
M$_\odot$.  We estimate that the chance probability of finding such a
galaxy cluster in the BAT error circle of \grb\ is about 0.1-1$\%$.  
This result highlights the ability to associate short GRBs with 
galaxy clusters based on $\gamma$-ray positions alone, thus removing a 
potential bias in favor of associations with gas-rich galaxies when 
relying on afterglow positions.

It has been suggested that the relative fraction of early- and
late-type host galaxies of short GRBs can be used to constrain the age
distribution of the progenitor population \citep{zr06}.  Making the
association between \grb\ and EDCC 493, and using all of the available
observations to date, we find that of the \swift\ short GRBs five
appear to be associated with early-type galaxies (050509b, 050724,
050813, 050911, 060502b) and two are associated with late-type
galaxies (050709, 051221a).  Taken at face value, this would argue for
an age distribution, $P(\tau)\propto \tau^n$ with $n\sim 2$
\citep{zr06}.  Naturally, in making a more accurate derivation of this
value, one has to take into account the respective probability of
association for each burst\footnotemark\footnotetext{For example, the
probability of association for GRB\,050509b is estimated at
3-4$\sigma$ \citep{gso+05,peh+05,bpp+06}, while that for GRB\,060502b
is only $2\sigma$ \citep{bpc+06}.}.  In fact if we consider only
secure associations, the relative numbers are $1\!:\!2$ instead of
$5\!:\!2$, leading to $n\sim -1$.

Independent of associations with individual galaxies, the fraction of
short GRBs in clusters is also of interest in assessing the age
distribution.  Of the sixteen available bursts, three have claimed
cluster associations\footnotemark\footnotetext{This fraction does not
change if we include the claimed cluster association for GRB\,790613
out of four IPN short burst error boxes searched by \citet{gno+05}.}
(050509b, 050813, 050911).  Within the uncertainty, this fraction is
in rough agreement with the value of $\sim 10\%$ for the overall
fraction of stellar mass in galaxy clusters \citep{fhp98}, or roughly
$20\%$ in clusters equal to or more massive than EDCC 493
\citep{ebc+05}.  Of course, not all stars are capable of producing
short GRBs, but assuming that the initial mass function and binary
fractions are independent of galaxy properties, the total stellar mass
provides a good proxy for the mass in short GRB progenitors.  Thus, at
the present there is reasonable agreement between the fraction of
short GRBs and the baseline fraction of stellar mass in galaxy
clusters.

Finally, we note that the frequency of short GRBs in galaxy clusters
may reflect a potential association with globular clusters.  The
latter are thought to provide an efficient environment for the
production of DNS binaries, and may account for a substantial fraction
of all short GRB progenitors \citep{gpm06}.  In particular, the
specific frequency\footnotemark\footnotetext{The specific frequency is
defined as the globular cluster number normalized to $M_V=-15$ mag
\citep{hb81}.} of globular clusters increases significantly from a
value of $\sim 1$ for S+Irr galaxies to $\sim 4$ for E+S0 galaxies,
and $\sim 12$ for cD galaxies \citep{har91}.  Given that massive
ellipticals are over-represented in galaxy clusters compared to the
field, we expect that an association with globular clusters will
increase the fraction of galaxy cluster associations compared to the
baseline level of $\sim 10-20\%$ indicated above.  Thus, continued
searches for galaxy clusters hosting short GRBs, and more detailed
predictions for the expected fraction as a function of cluster mass
and redshift, may hold the key to a clearer understanding of the
progenitor population.

\acknowledgements 

We thank A.~Dressler, A.~Gal-Yam, M.~Gladders, D.~Kawata, 
F.~Schweizer, and A.~Soderberg for helpful discussions.  E.B.~is 
supported by NASA through Hubble Fellowship grant HST-01171.01 awarded 
by the Space Telescope Science Institute, which is operated by AURA, 
Inc.~for NASA under contract NAS 5-26555.  M.-S.~S.~acknowledges 
support from the Observatories of the Carnegie Institution of 
Washington and Korean Science and Engineering Foundation Grant 
KOSEF-2005-215-C00056 funded by the Korean government (MOST).


\begin{thebibliography}{}

\bibitem[\protect\citeauthoryear{{Arnaud}, {Pointecouteau}, \&
  {Pratt}}{{Arnaud} et~al.}{2005}]{app05}
{Arnaud}, M., {Pointecouteau}, E.,  \& {Pratt}, G.~W. 2005, \aap, 441, 893

\bibitem[\protect\citeauthoryear{{Baldry} et~al.}{{Baldry}
  et~al.}{2006}]{bbb+06}
{Baldry}, I., {Balogh}, M., {Bower}, R., {Glazebrook}, K., {Nichol}, R.,
  {Bamford}, S.,  \& {Budavari}, T. 2006, ArXiv Astrophysics e-prints, 
  astro-ph/0607648

\bibitem[\protect\citeauthoryear{{Barthelmy} et~al.}{{Barthelmy}
  et~al.}{2005}]{bcb+05}
{Barthelmy}, S.~D., et~al. 2005, \nat, 438, 994

\bibitem[\protect\citeauthoryear{{Beers}, {Flynn}, \& {Gebhardt}}{{Beers}
  et~al.}{1990}]{bfg90}
{Beers}, T.~C., {Flynn}, K.,  \& {Gebhardt}, K. 1990, \aj, 100, 32

\bibitem[\protect\citeauthoryear{{Berger}}{{Berger}}{2005}]{ber05}
{Berger}, E. 2005, GRB Coordinates Network, 3962, 1

\bibitem[\protect\citeauthoryear{{Berger}}{{Berger}}{2006}]{ber06}
{Berger}, E. 2006, in AIP Conf. Proc. 838: Gamma-Ray Bursts in the Swift Era,
  ed. S.~S. {Holt}, N.~{Gehrels}, \& J.~A. {Nousek}, 33

\bibitem[\protect\citeauthoryear{{Berger} et~al.}{{Berger}
  et~al.}{2005}]{bpc+05}
{Berger}, E., et~al. 2005, \nat, 438, 988

\bibitem[\protect\citeauthoryear{{Bernardi} et~al.}{{Bernardi}
  et~al.}{1998}]{brc+98}
{Bernardi}, M., {Renzini}, A., {da Costa}, L.~N., {Wegner}, G., {Alonso},
  M.~V., {Pellegrini}, P.~S., {Rit{\'e}}, C.,  \& {Willmer}, C.~N.~A. 1998,
  \apjl, 508, L143

\bibitem[\protect\citeauthoryear{{Bertin} \& {Arnouts}}{{Bertin} \&
  {Arnouts}}{1996}]{ba96}
{Bertin}, E.,  \& {Arnouts}, S. 1996, \aaps, 117, 393

\bibitem[\protect\citeauthoryear{{Bloom} et~al.}{{Bloom}
  et~al.}{2006a}]{bpc+06}
{Bloom}, J.~S., et~al. 2006a, ArXiv Astrophysics e-prints, 
astro-ph/0607223

\bibitem[\protect\citeauthoryear{{Bloom} et~al.}{{Bloom}
  et~al.}{2006b}]{bpp+06}
{Bloom}, J.~S., et~al. 2006b, \apj, 638, 354

\bibitem[\protect\citeauthoryear{{Burrows} et~al.}{{Burrows}
  et~al.}{2006}]{bgc+06}
{Burrows}, D.~N., et~al. 2006, ArXiv Astrophysics e-prints, 
astro-ph/0604320

\bibitem[\protect\citeauthoryear{{Campana} et~al.}{{Campana}
  et~al.}{2006}]{ctl+06}
{Campana}, S., et~al. 2006, \aap, 454, 113

\bibitem[\protect\citeauthoryear{{Champion} et~al.}{{Champion}
  et~al.}{2004}]{clm+04}
{Champion}, D.~J., {Lorimer}, D.~R., {McLaughlin}, M.~A., {Cordes}, J.~M.,
  {Arzoumanian}, Z., {Weisberg}, J.~M.,  \& {Taylor}, J.~H. 2004, \mnras, 350,
  L61

\bibitem[\protect\citeauthoryear{{Cole} et~al.}{{Cole} et~al.}{2001}]{cnb+01}
{Cole}, S., et~al. 2001, \mnras, 326, 255

\bibitem[\protect\citeauthoryear{Cucchiara et~al.}{Cucchiara
  et~al.}{2006}]{cfb+06}
Cucchiara, A., Fox, D.~B., Berger, E.,  \& Price, P.~A. 2006, GRB Coordinates
  Network, 5470, 1

\bibitem[\protect\citeauthoryear{{de Luca} et~al.}{{de Luca}
  et~al.}{2005}]{gcn4274}
{de Luca}, A., {Caraveo}, P., {Esposito}, P., {Mereghetti}, S.,  \& {Tiengo},
  A. 2005, GRB Coordinates Network, 4274, 1

\bibitem[\protect\citeauthoryear{{De Lucia} et~al.}{{De Lucia}
  et~al.}{2006}]{dsw+06}
{De Lucia}, G., {Springel}, V., {White}, S.~D.~M., {Croton}, D.,  \&
  {Kauffmann}, G. 2006, \mnras, 366, 499

\bibitem[\protect\citeauthoryear{{Dickey} \& {Lockman}}{{Dickey} \&
  {Lockman}}{1990}]{dl90}
{Dickey}, J.~M.,  \& {Lockman}, F.~J. 1990, \araa, 28, 215

\bibitem[\protect\citeauthoryear{{Dressler}}{{Dressler}}{1980}]{dre80}
{Dressler}, A. 1980, \apj, 236, 351

\bibitem[\protect\citeauthoryear{{Eichler} et~al.}{{Eichler}
  et~al.}{1989}]{elp+89}
{Eichler}, D., {Livio}, M., {Piran}, T.,  \& {Schramm}, D.~N. 1989, \nat, 340,
  126

\bibitem[\protect\citeauthoryear{{Eke} et~al.}{{Eke} et~al.}{2005}]{ebc+05}
{Eke}, V.~R., {Baugh}, C.~M., {Cole}, S., {Frenk}, C.~S., {King}, H.~M.,  \&
  {Peacock}, J.~A. 2005, \mnras, 362, 1233

\bibitem[\protect\citeauthoryear{{Fox} et~al.}{{Fox} et~al.}{2005}]{ffp+05}
{Fox}, D.~B., et~al. 2005, \nat, 437, 845

\bibitem[\protect\citeauthoryear{{Fukugita}, {Hogan}, \& {Peebles}}{{Fukugita}
  et~al.}{1998}]{fhp98}
{Fukugita}, M., {Hogan}, C.~J.,  \& {Peebles}, P.~J.~E. 1998, \apj, 503, 518

\bibitem[\protect\citeauthoryear{{Gal-Yam} et~al.}{{Gal-Yam}
  et~al.}{2005}]{gno+05}
{Gal-Yam}, A., et~al. 2005, ArXiv Astrophysics e-prints, 
astro-ph/0509891

\bibitem[\protect\citeauthoryear{{Gehrels} et~al.}{{Gehrels}
  et~al.}{2005}]{gso+05}
{Gehrels}, N., et~al. 2005, \nat, 437, 851

\bibitem[\protect\citeauthoryear{{Grindlay}, {Portegies Zwart}, \&
  {McMillan}}{{Grindlay} et~al.}{2006}]{gpm06}
{Grindlay}, J., {Portegies Zwart}, S.,  \& {McMillan}, S. 2006, Nature Physics,
  2, 116

\bibitem[\protect\citeauthoryear{{Guetta} \& {Piran}}{{Guetta} \&
  {Piran}}{2006}]{gp06}
{Guetta}, D.,  \& {Piran}, T. 2006, \aap, 453, 823

\bibitem[\protect\citeauthoryear{{Harris}}{{Harris}}{1991}]{har91}
{Harris}, W.~E. 1991, \araa, 29, 543

\bibitem[\protect\citeauthoryear{{Harris} \& {van den Bergh}}{{Harris} \& {van
  den Bergh}}{1981}]{hb81}
{Harris}, W.~E.,  \& {van den Bergh}, S. 1981, \aj, 86, 1627

\bibitem[\protect\citeauthoryear{{Hjorth} et~al.}{{Hjorth}
  et~al.}{2005a}]{hsg+05}
{Hjorth}, J., et~al. 2005a, \apjl, 630, L117

\bibitem[\protect\citeauthoryear{{Hjorth} et~al.}{{Hjorth}
  et~al.}{2005b}]{hwf+05}
{Hjorth}, J., et~al. 2005b, \nat, 437, 859

\bibitem[\protect\citeauthoryear{{Holland} et~al.}{{Holland}
  et~al.}{2005}]{gcn4034}
{Holland}, S.~T., {Barthelmy}, S., {Beardmore}, A., {Gehrels}, N., {Kennea},
  J., {Page}, K., {Palmer}, D.,  \& {Rosen}, S. 2005, GRB Coordinates Network,
  4034, 1

\bibitem[\protect\citeauthoryear{{Horner}, {Mushotzky}, \& {Scharf}}{{Horner}
  et~al.}{1999}]{hms99}
{Horner}, D.~J., {Mushotzky}, R.~F.,  \& {Scharf}, C.~A. 1999, \apj, 520, 78

\bibitem[\protect\citeauthoryear{{Kulkarni}}{{Kulkarni}}{2005}]{kul05}
{Kulkarni}, S.~R. 2005, ArXiv Astrophysics e-prints, astro-ph/0510256

\bibitem[\protect\citeauthoryear{{Kuntschner} et~al.}{{Kuntschner}
  et~al.}{2002}]{ksc+02}
{Kuntschner}, H., {Smith}, R.~J., {Colless}, M., {Davies}, R.~L., {Kaldare},
  R.,  \& {Vazdekis}, A. 2002, \mnras, 337, 172

\bibitem[\protect\citeauthoryear{{Li} \& {Paczy{\'n}ski}}{{Li} \&
  {Paczy{\'n}ski}}{1998}]{lp98}
{Li}, L.-X.,  \& {Paczy{\'n}ski}, B. 1998, \apjl, 507, L59

\bibitem[\protect\citeauthoryear{{Lumsden} et~al.}{{Lumsden}
  et~al.}{1992}]{lnc+92}
{Lumsden}, S.~L., {Nichol}, R.~C., {Collins}, C.~A.,  \& {Guzzo}, L. 1992,
  \mnras, 258, 1

\bibitem[\protect\citeauthoryear{{Mulchaey} et~al.}{{Mulchaey}
  et~al.}{2006}]{mlf+06}
{Mulchaey}, J.~S., {Lubin}, L.~M., {Fassnacht}, C., {Rosati}, P.,  \&
  {Jeltema}, T.~E. 2006, \apj, 646, 133

\bibitem[\protect\citeauthoryear{{Mushotzky} \& {Loewenstein}}{{Mushotzky} \&
  {Loewenstein}}{1997}]{ml97}
{Mushotzky}, R.~F.,  \& {Loewenstein}, M. 1997, \apjl, 481, L63

\bibitem[\protect\citeauthoryear{{Nakar}, {Gal-Yam}, \& {Fox}}{{Nakar}
  et~al.}{2005}]{ngf05}
{Nakar}, E., {Gal-Yam}, A.,  \& {Fox}, D.~B. 2005, ArXiv Astrophysics 
e-prints, astro-ph/0511254

\bibitem[\protect\citeauthoryear{{Narayan}, {Paczynski}, \& {Piran}}{{Narayan}
  et~al.}{1992}]{npp92}
{Narayan}, R., {Paczynski}, B.,  \& {Piran}, T. 1992, \apjl, 395, L83

\bibitem[\protect\citeauthoryear{{Page} et~al.}{{Page} et~al.}{2006}]{pkl+06}
{Page}, K.~L., et~al. 2006, \apjl, 637, L13

\bibitem[\protect\citeauthoryear{{Pedersen} et~al.}{{Pedersen}
  et~al.}{2005}]{peh+05}
{Pedersen}, K., et~al. 2005, \apjl, 634, L17

\bibitem[\protect\citeauthoryear{{Prochaska} et~al.}{{Prochaska}
  et~al.}{2006}]{pbc+06}
{Prochaska}, J.~X., et~al. 2006, \apj, 642, 989

\bibitem[\protect\citeauthoryear{{Rosati} et~al.}{{Rosati}
  et~al.}{1998}]{rcn+98}
{Rosati}, P., {della Ceca}, R., {Norman}, C.,  \& {Giacconi}, R. 1998, \apjl,
  492, L21

\bibitem[\protect\citeauthoryear{{Rosswog}, {Ramirez-Ruiz}, \&
  {Davies}}{{Rosswog} et~al.}{2003}]{rrd03}
{Rosswog}, S., {Ramirez-Ruiz}, E.,  \& {Davies}, M.~B. 2003, \mnras, 345, 1077

\bibitem[\protect\citeauthoryear{{Soderberg} et~al.}{{Soderberg}
  et~al.}{2006}]{sbk+06}
{Soderberg}, A.~M., et~al. 2006, ArXiv Astrophysics e-prints, 
astro-ph/0601455

\bibitem[\protect\citeauthoryear{{Thomas} et~al.}{{Thomas}
  et~al.}{2005}]{tmb+05}
{Thomas}, D., {Maraston}, C., {Bender}, R.,  \& {Mendes de Oliveira}, C. 2005,
  \apj, 621, 673

\bibitem[\protect\citeauthoryear{{Villasenor} et~al.}{{Villasenor}
  et~al.}{2005}]{vlr+05}
{Villasenor}, J.~S., et~al. 2005, \nat, 437, 855

\bibitem[\protect\citeauthoryear{{White} \& {Springel}}{{White} \&
  {Springel}}{2000}]{ws00}
{White}, S.~D.~M.,  \& {Springel}, V. 2000, in The First Stars: Proceedings of
  the MPA/ESO Workshop Held at Garching, Germany, 4-6 August 1999, ESO
  ASTROPHYSICS SYMPOSIA. ISBN 3-540-67222-2. Edited by A. Weiss, T.G. Abel, and
  V. Hill. Springer-Verlag, 2000, p. 327, ed. A.~{Weiss}, T.~G. {Abel}, \&
  V.~{Hill}, 327

\bibitem[\protect\citeauthoryear{{Whitmore}, {Gilmore}, \& {Jones}}{{Whitmore}
  et~al.}{1993}]{wgj93}
{Whitmore}, B.~C., {Gilmore}, D.~M.,  \& {Jones}, C. 1993, \apj, 407, 489

\bibitem[\protect\citeauthoryear{{Zheng} \& {Ramirez-Ruiz}}{{Zheng} \&
  {Ramirez-Ruiz}}{2006}]{zr06}
{Zheng}, Z.,  \& {Ramirez-Ruiz}, E. 2006, ArXiv Astrophysics e-prints, 
astro-ph/0601622

\end{thebibliography}

\clearpage
\begin{deluxetable}{llcccclcc}
\tablecolumns{9}
\tabcolsep0.05in\footnotesize
\tablewidth{0pc}
\tablecaption{Journal of Spectroscopic Observations
\label{tab:obs}}
\tablehead {
\colhead {GRB}            &
\colhead {Date}           &
\colhead {Grism}          &
\colhead {Filter}         &
\colhead {Exp.~Time}      &
\colhead {Airmass}        &
\colhead {$\lambda$}      &         
\colhead {Objects}        &              
\colhead {Redshifts}      \\
\colhead {  }             &
\colhead {(UT)}           &
\colhead {(l mm$^{-1}$)}  &
\colhead {  }             &
\colhead {(s)}            &
\colhead {  }             &
\colhead {($\mu$m)}       &         
\colhead {  }             &              
\colhead {  }                    
}
\startdata
050709  & 2006 June 28.23 & 300  & OG590       & 4800 & 1.47 & $0.58-1.00$     & 15 & 10 \\
        & 2006 June 28.30 & 300  & OG590       & 4800 & 1.12 & $0.58-1.00$     & 11 & 5  \\
        & 2006 June 28.39 & 300  & OG590       & 4000 & 1.02 & $0.58-1.00$     & 7  & 6  \\
050724  & 2006 June 28.96 & 300  & W4800-7800  & 4800 & 1.47 & $0.48-0.78$     & 21 & 18 \\
        & 2006 June 29.02 & 1090 & None        & 3600 & 1.13 & $0.39-0.68\,^a$ & 21 & 19 \\
051221a & 2006 June 29.28 & 300  & W4800-7800  & 6000 & 1.59 & $0.48-0.78$     & 41 & 34 \\
        & 2006 June 29.36 & 300  & W4800-7800  & 6900 & 1.44 & $0.48-0.78$     & 38 & 27 \\
050911  & 2006 June 30.28 & 300  & None        & 6000 & 1.70 & $0.40-1.00$     & 25 & 21 \\
        & 2006 June 30.36 & 300  & None        & 5800 & 1.18 & $0.40-1.00$     & 13 & 11
\enddata
\tablecomments{Observing log for the four short GRB fields  
targeted on the nights of 2006 June 28--30 UT.  $^a$ The 
wavelength coverage depends on the location of the slit, 
ranging from $0.36-0.62$ $\mu$m to $0.46-0.74$ $\mu$m.}
\end{deluxetable}

\clearpage
\begin{deluxetable}{lllcccccc}
\tablecolumns{8}
\tabcolsep0.05in\footnotesize
\tablewidth{0pc}
\tablecaption{X-ray Observations of Short GRB Fields
\label{tab:xray}}
\tablehead {
\colhead {GRB}                      &
\colhead {$z$}                      &
\colhead {Date}                     &
\colhead {Telescope}                &
\colhead {Exp.~Time}                &
\colhead {$N({\rm HI})\,^a$}        &
\colhead {Count Rate$^{b}$}         &
\colhead {$F_X\,^c$}                &
\colhead {$L_X$}                    \\
\colhead {}                         &
\colhead {}                         &
\colhead {(UT)}                     &
\colhead {}                         &
\colhead {(s)}                      &
\colhead {($10^{20}$ cm$^{-2}$)}    &
\colhead {(s$^{-1}$)}               &
\colhead {($10^{-14}$ erg s$^{-1}$ cm$^{-2}$)} &                
\colhead {($10^{42}$ erg s$^{-1}$)}                
}
\startdata
050509b & 0.226  & 2005 May 9.17   & XRT & $32080\,^d$  & 1.27 & 0.0092 & $37^{+6}_{-5}$ & $45^{+7}_{-6}$ \\
050709	& 0.161  & 2005 Jul.~11.55 & XRT & 17316 & \nod & \nod      & \nod   & \nod      \\
        &        & 2005 Jul.~12.43 & XRT & 4977  & \nod & \nod      & \nod   & \nod      \\
        &        & 2005 Jul.~13.11 & XRT & 7035  & 1.23 & $<0.0010$ & $<2.7$ & $<1.6$    \\
        &        & 2005 Jul.~25.86 & CXO & 18284 & \nod & $<0.0094$ & $<3.8$ & $<2.3$    \\
050724	& 0.257  & 2005 Jul.~29.02 & XRT & 24880 & \nod & \nod      & \nod   & \nod      \\
        &        & 2005 Jul.~30.02 & XRT & 21128 & 14.5 & $<0.0010$ & $<3.8$ & $<6.0$    \\
        &        & 2005 Jul.~26.84 & CXO & 49955 & \nod & $<0.0064$ & $<3.6$ & $<5.7$    \\ 
050813	& 1.8?   & 2005 Aug.~13.28 & XRT & 14210 & \nod & \nod      & \nod   & \nod      \\
      	&        & 2005 Aug.~16.03 & XRT & 11438 & \nod & \nod      & \nod   & \nod      \\
      	&        & 2005 Aug.~19.04 & XRT & 9265  & \nod & \nod      & \nod   & \nod      \\
      	&        & 2005 Aug.~21.65 & XRT & 13985 & 4.08 & $<0.0011$ & $<3.3$ & $<270$    \\
050906	& \nod   & 2005 Sep.~6.44  & XRT & 5687  & 5.60 & $<0.0024$ & $<7.4$ & \nod      \\  
050911	& 0.165  & 2005 Sep.~11.86 & XRT & 6225  & \nod & \nod      & \nod   & \nod      \\
        &        & 2005 Sep.~12.59 & XRT & 9755  & \nod & \nod      & \nod   & \nod      \\
        &        & 2005 Sep.~14.06 & XRT & 12983 & \nod & \nod      & \nod   & \nod      \\
        &        & 2005 Sep.~18.27 & XRT & 16723 & 2.70 & 0.0033 & $7.7^{+2.0}_{-1.9}$ & $4.9^{+1.3}_{-1.2}$ \\
050925$\,^e$ & \nod & 2005 Sep.~25.38 & XRT & 37152 & \nod & \nod   & \nod   & \nod      \\
        &        & 2006 Apr.~6.03  & XRT & 14841 & \nod & \nod      & \nod   & \nod      \\
        &        & 2006 May 8.76   & XRT & 8735  & 114  & $<0.0009$ & $<13$  & \nod      \\
051105a	& \nod   & 2005 Nov.~5.27  & XRT & 55846 & 2.89 & $<0.0007$ & $<2.1$ & \nod      \\
051114	& \nod   & 2005 Nov.~15.64 & XRT & 4750  & \nod &           & \nod   & \nod      \\
        &        & 2005 Nov.~15.91 & XRT & 10753 & 1.66 & $<0.0011$ & $<3.0$ & \nod      \\ 
051210	& \nod   & 2005 Dec.~10.24 & XRT & 37401 & 2.14 & $<0.0009$ & $<2.6$ & \nod      \\
051221a	& 0.546  & 2006 Jan.~2.09  & XRT & 56663 & 6.58 & $<0.0007$ & $<2.2$ & $<17$     \\       
051227	& \nod   & 2005 Dec.~30.03 & XRT & 45913 & \nod & \nod      & \nod   & \nod      \\
        &        & 2006 Jan.~1.02  & XRT & 10216 & \nod & \nod      & \nod   & \nod      \\
        &        & 2006 Jan.~2.11  & XRT & 27203 & 4.20 & $<0.0006$ & $<1.7$ & \nod      \\
060121	& \nod   & 2006 Jan.~28.01 & XRT & 25272 & \nod & \nod      & \nod   & \nod      \\
        &        & 2006 Feb.~2.04  & XRT & 18420 & 1.66 & $<0.0007$ & $<2.0$ & \nod      \\
060313	& \nod   & 2006 Mar.~15.00 & XRT & 40463 & 4.65 & $<0.0008$ & $<2.5$ & \nod      \\
060502b & 0.287? & 2006 May 2.73   & XRT & 29630 & 4.32 & $<0.0010$ & $<3.0$ & $<6.0$    \\       
060801  & 1.131  & 2006 Aug.~2.04  & XRT & 20942 & \nod & \nod      & \nod   & \nod      \\
        &        & 2006 Aug.~3.05  & XRT & 38051 & 1.54 & $<0.0007$ & $<2.0$ & $<68$
\enddata
\tablecomments{X-ray observations of short GRB fields obtained 
with \swift/XRT and {\it Chandra}.  $^a$ Galactic neutral 
hydrogen column density from \citet{dl90}.  $^b$  Upper limits 
are calculated as $3\sigma$ of the number of counts in a $2'$ 
radius aperture, corresponding to $\sim 400-700$ kpc at $z\sim 
0.2-0.5$.  For multiple observations we provide the upper limit 
based on the combined data.  $^c$ Unabsorbed flux.  Conversion 
from count rate to flux assumes a thermal bremsstrahlung model 
with $kT=1$ keV and the Galactic absorbing column densities 
given in column 6.  $^d$ First 0.5 hr of data removed to 
eliminate the contribution from the afterglow.  $^e$ This 
object is possibly a Galactic soft $\gamma$-ray repeater 
\citep{gcn4034}.  No diffuse X-ray emission is evident in 
XMM-{\it Newton} observations of this field \citep{gcn4274}.  
References for redshifts (in order): 
\citet{bpp+06}, \citet{ffp+05}, \citet{bpc+05}, \citet{ber06}, 
\citet{sbk+06}, \citet{bpc+06}, and \citet{cfb+06}.}
\end{deluxetable}

\clearpage
\begin{deluxetable}{llllc}
\tablecolumns{5}
\tabcolsep0.2in\footnotesize
\tablewidth{0pc}
\tablecaption{Properties of Spectroscopically-confirmed EDCC 493 Cluster Members
\label{tab:050911}}
\tablehead {
\colhead {RA}         &
\colhead {Dec}        &
\colhead {$I$}        &
\colhead {$z$}        &
\colhead {Type}       \\
\colhead {(J2000)}    &
\colhead {(J2000)}    &
\colhead {(mag)}      &
\colhead {}           &            
\colhead {}                       
}
\startdata
\ra{00}{54}{52.579} & \dec{-38}{51}{41.27} & 20.48 & $0.1615\pm 0.0010$ & abs+em \\
\ra{00}{54}{56.067} & \dec{-38}{54}{27.59} & 20.76 & $0.1649\pm 0.0008$ & abs+em \\   
\ra{00}{54}{56.472} & \dec{-38}{51}{49.61} & 21.62 & $0.1642\pm 0.0012$ & abs    \\
\ra{00}{54}{56.861} & \dec{-38}{49}{43.67} & 20.00 & $0.1655\pm 0.0011$ & abs    \\
\ra{00}{54}{58.484} & \dec{-38}{51}{58.70} & 20.36 & $0.1627\pm 0.0003$ & abs    \\
\ra{00}{54}{58.965} & \dec{-38}{50}{56.07} & 20.36 & $0.1651\pm 0.0008$ & abs    \\
\ra{00}{55}{00.664} & \dec{-38}{52}{30.06} & 17.57 & $0.1649\pm 0.0005$ & abs    \\
\ra{00}{55}{01.482} & \dec{-38}{54}{13.05} & 20.53 & $0.1615\pm 0.0005$ & abs+em \\
\ra{00}{55}{03.433} & \dec{-38}{49}{22.48} & 18.98 & $0.1649\pm 0.0014$ & abs    \\
\ra{00}{55}{04.740} & \dec{-38}{50}{42.18} & 18.90 & $0.1640\pm 0.0005$ & abs    \\
\ra{00}{55}{05.205} & \dec{-38}{52}{37.09} & 19.08 & $0.1673\pm 0.0018$ & abs    \\
\ra{00}{55}{05.679} & \dec{-38}{51}{00.31} & 21.14 & $0.1673\pm 0.0005$ & abs    \\
\ra{00}{55}{11.168} & \dec{-38}{52}{43.84} & 20.80 & $0.1615\pm 0.0010$ & abs    \\
\ra{00}{55}{11.820} & \dec{-38}{52}{30.59} & 20.03 & $0.1656\pm 0.0006$ & abs  
\enddata
\tablecomments{Positions and observed properties of the EDCC 
493 cluster members confirmed in our observations.  The last 
column indicates the redshift type.}
\end{deluxetable}

\clearpage
\begin{figure}
\epsscale{0.8}
\caption{HST/ACS F814W image of the field of GRB\,050709 showing our
spectroscopic targets and their measured redshifts.  The host galaxy
is marked with a thick black circle.  Question marks indicate targets
for which a redshift could not be determined.
\label{fig:050709}}
\end{figure}

\clearpage
\begin{figure}
\epsscale{0.8}
\caption{LDSS3 $I$-band image of the field of GRB\,050724 showing our
spectroscopic targets and their measured redshifts.  The host galaxy
is marked with a thick black circle.  Question marks indicate targets
for which a redshift could not be determined.
\label{fig:050724}}
\end{figure}

\clearpage
\begin{figure}
\epsscale{0.8}
\caption{LCO/du Pont $I$-band image of the field of \grb\ showing our
spectroscopic targets and their measured redshifts.  The BAT $2.8'$
radius error circle is marked by a thick black circle.  The cluster
EDCC 493 is clearly visible in the southeast corner of the error
circle.  Question marks indicate targets for which a redshift could 
not be determined.
\label{fig:050911}}
\end{figure}

\clearpage
\begin{figure}
\epsscale{0.8}
\caption{Gemini/GMOS $r'$-band image of the field of GRB\,051221a
showing our spectroscopic targets and their measured redshifts.  The
host galaxy (not targeted in our observations) is marked with an
arrow.  Question marks indicate targets for which a redshift could 
not be determined.
\label{fig:051221a}}
\end{figure}

\clearpage

\clearpage

\clearpage

\clearpage

\clearpage
\begin{figure}
\epsscale{1}
\includegraphics[angle=270,width=7.5in]{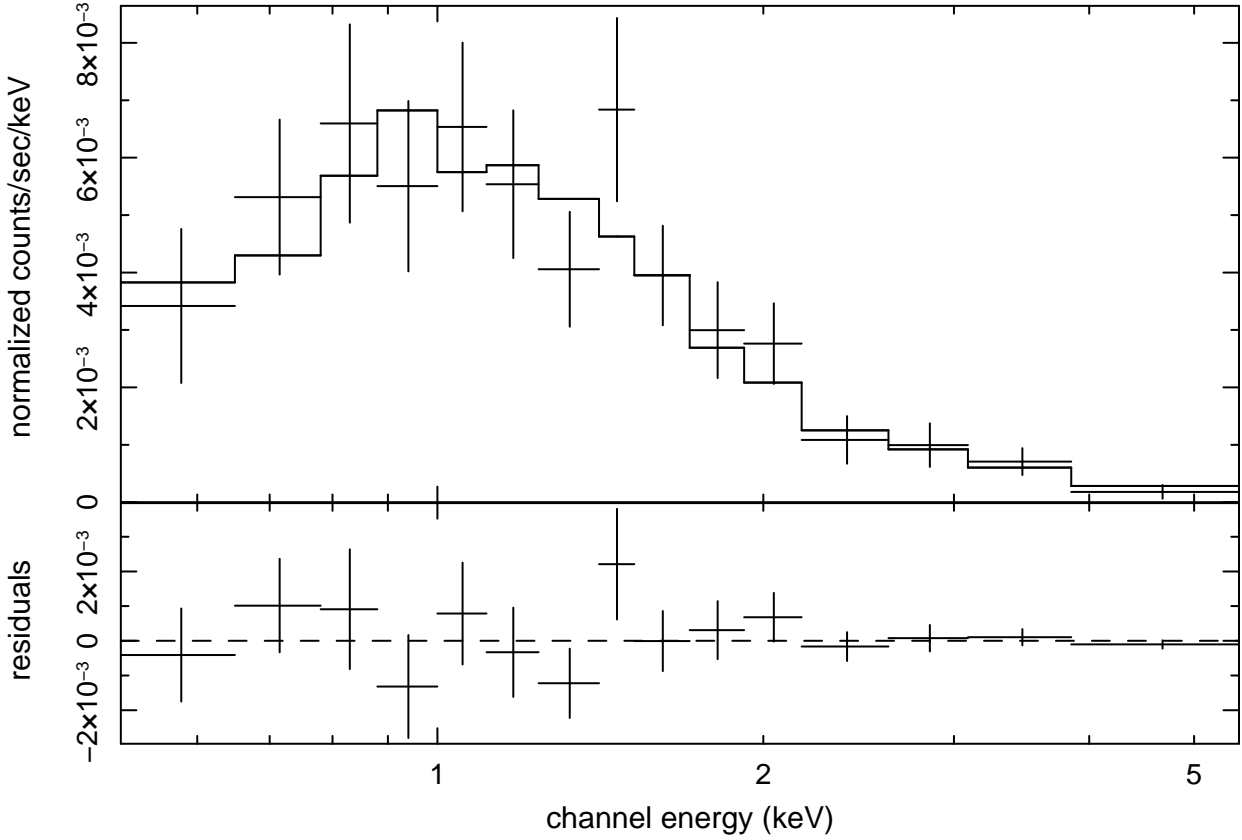}   
\caption{Spectrum of the diffuse X-ray emission from the cluster ZwCl
1234.0+02916 coincident with GRB\,050509b as observed by XRT.  The
data are binned with at least 25 counts per bin.  We use a MEKAL
model fit with an abundance fixed at $0.3$ Z$_\odot$.  The best-fit
parameters are $kT=3.0^{+0.9}_{-0.6}$ keV, an unabsorbed luminosity
of $L_X=4.5^{+0.7}_{-0.6}\times 10^{43}$ erg s$^{-1}$, and an
absorbing column density of $N_H=1.1^{+0.6}_{-0.5}\times 10^{21}$
cm$^{-2}$ ($\chi^2_r=0.64$ for 12 degrees of freedom).
\label{fig:g050509bspec}} 
\end{figure}

\end{document}